\documentstyle[12pt]{article}
\advance\textheight by \topskip
\begin{document}
\begin{flushright}
SU-ITP-95-14\\
hep-th/9507022\\
July 3, 1995\\
\end{flushright}
\vspace{-0.2cm}
\begin{center}
\baselineskip=16pt

{\Large\bf  EXACT SUPERSYMMETRIC  MASSIVE  \\
\vskip 0.6 cm
AND MASSLESS WHITE HOLES}\\

\vskip 1.5cm

{\bf Renata Kallosh}\footnote {E-mail:
kallosh@physics.stanford.edu} {\bf ~ and~~ Andrei Linde}\footnote{E-mail:
linde@physics.stanford.edu}
 \vskip 0.05cm
Physics Department, Stanford University, Stanford   CA 94305\\
\vskip 0.7 cm

\end{center}
\vskip 1 cm
\centerline{\bf ABSTRACT}
\begin{quotation}

We study special points in the moduli space of vacua at which supersymmetric
electric solutions of the heterotic string theory become massless. We
concentrate on configurations for which supersymmetric non-renormalization
theorem is valid. Those are ten-dimensional supersymmetric string waves and
generalized fundamental strings with  $SO(8)$  holonomy group. From these we
find the four dimensional spherically symmetric configurations which saturate
the BPS bound, in particular near the points of the vanishing ADM mass. The
non-trivial massless supersymmetric states in this class exist only in the
presence of non-Abelian vector fields.

We also find a new class of supersymmetric massive solutions, closely related
to the massless ones. A distinctive property of all these objects, either
massless or massive, is the existence of gravitational repulsion. They reflect
all particles with nonvanishing mass and/or angular momentum, and therefore
they can be called white holes (repulsons), in contrast to black holes which
tend to absorb
particles of all kinds. If such objects can exist, we will have the first
realization of  the universal gravitational force which repels all  particles
with the strength proportional to their mass and therefore can be associated
with antigravity.

\end{quotation}
\newpage
\baselineskip=16pt

\section{Introduction}

 The purpose of this paper is to exhibit some peculiar features of  exact
supersymmetric solutions of the heterotic string theory. We will consider the
most unusual properties of these solutions which all saturate the
supersymmetric positivity bound in the limit when the mass of such
configurations tends to zero,
\begin{equation}
M^2 = Z^A {\cal R } _{AB}(\Phi_0) Z^B \rightarrow 0 \ .
\end{equation}
Here ${\cal R} $ is a    continuous function of the asymptotic values
of the  scalar fields $\Phi_0$ and $Z$ are electric and magnetic charges.
It has been pointed out recently by Hull and Townsend \cite{HullT} and  Witten
\cite{W} and Strominger \cite{Strom} that  since the matrix ${\cal R} $ is a
continuous function of  $\Phi_0$,
the masses of the Bogomolny states are also continuous functions of  $\Phi_0$.
In particular, for some values of  $\Phi_0$  massless states may exist
which saturate the bound.

The first explicit example of such a state was given by Behrndt \cite{Klaus2}
and it was interpreted as a massless black hole. This was a significant
progress. A wider class of similar solutions was obtained in
\cite{BHGAUGE}. The solutions were interpreted as $N_L=0$ states of the
toroidally compactified heterotic string. However, all these solutions did not
include non-Abelian fields which are necessary to cancel anomalies of
supersymmetry. Therefore it was not quite clear whether these solutions survive
and remain massless with an account taken of $\alpha'$ corrections. The
importance of quantum corrections to supersymmetry transformations
increases  in the situation when it is known that the massless states  may
present the
points of enhanced gauge symmetry.   Anomalies of supersymmetry result from the
Lorentz anomaly in the effective heterotic string action and may spoil all
conjectures about the exactness of the BPS bound. The anomalies can be cured
when
Yang-Mills fields  are included according to the Green-Schwarz mechanism of
cancellation of anomalies.

 In the present paper we will consider a
general class of supersymmetric solutions including  non-Abelian fields. We
will
find massless states which may correspond either to  black holes  or to waves,
and
which remain exact solutions of equations of motion due to the presence of
non-Abelian fields even with an account taken of
$\alpha'$ corrections. In other words,
we will find massless BPS states  which   saturate
the supersymmetric positivity bound and which are free of
anomalies of supersymmetry.

We will find also a  class of anomaly free massive configurations closely
related to the massless black holes. A rather unusual property of all these
configurations
(either massive or massless) is that instead of the  usual black hole horizon,
which
{\it absorbs} all particles falling into the black hole, they  have a repulsive
(i.e. antigravitating) naked singularity
which {\it reflects} all test particles. Since the totally reflecting surface
is not
black but white, it is more proper to call these new singular configurations
supersymmetric  white
holes,  or repulsons.\footnote{The name ``white holes'' referring to the
complete reflection  (as opposite to the complete absorbtion by black holes)
seems to be most adequate. Unfortunately,   many years ago this name  was used
for   hypothetical
objects  associated with the   time reversal of the gravitational
collapse. Physical relevance of such objects is rather doubtful
\cite{NovFrol}. Therefore we believe that that the name ``white holes'' is
essentially vacant and can be used    for the  repulsive singular
configurations   discussed in our paper. However, in order to completely avoid
collisions with  old terminology, we will often call our  massive and massless
white hole configurations ``repulsons.''  }
 The solutions can be obtained either directly in four-dimensional space or by
dimensional reduction of ten-dimensional gravitational waves. The  repulsive
singularity discussed above   is not present in the
ten-dimensional non-compactified version of the solution. It appears after the
compactification in those places of the four-dimensional space  where   the
volume of the six-dimensional compactified space shrinks to zero.

Our basic strategy in looking for massless states   is the following.
If a configuration  has one-half of unbroken
$N=4$ supersymmetry and describes a solutions of $N=4$ supergravity interacting
with some Abelian and non-Abelian vector multiplets,
the vanishing ADM mass simultaneously means the vanishing dilaton
charge\footnote{For solutions without the fundamental axion charge this can be
shown using the supersymmetry rules.}:
\begin{equation}
M \rightarrow 0 \hskip 1 cm  \Longleftrightarrow   \hskip 1 cm  \Sigma
\rightarrow 0 \ .
\end{equation}
 If we know any   ten-dimensional solution with one-half of unbroken  $N=1$
supersymmetry, we may use the fact that the corresponding four-dimensional
dilaton $e^{2 \phi}$ is related to the ten-dimensional dilaton $e^{2\hat \phi}$
as
follows:
\begin{equation}
e^{- 2 \phi} =  e^{- 2\hat \phi} \sqrt {\det G} \ ,
\end{equation}
where the matrix $ G $ describes the geometry of the internal six-dimensional
space.  Knowledge of the ten-dimensional solutions means that both $e^{- 2\hat
\phi}$  and
${\det G}$ are known.
The massless four-dimensional configurations saturating the BPS bound   are the
ones
in which the ten-dimensional dilaton charge $\hat \Sigma$ is  compensated by
the
modulus field  charge
$\sigma$, where  we define
\begin{eqnarray}
e^{- 2 \phi} &=& e^{- 2 \phi_0} + {\Sigma\over r} +\dots ,\nonumber\\
e^{- 2 \hat \phi} &=& e^{- 2\hat  \phi_0} + {\hat\Sigma\over r} +\dots ,
\nonumber\\
\sqrt {\det G} &=&( \sqrt {\det G})_0 +  {\sigma\over r} +\dots
\end{eqnarray}
The BPS-state is massless  when
\begin{equation}
( \sqrt {\det G})_0 \,\hat\Sigma + e^{- 2\hat  \phi_0} \,  \sigma   =0 \ .
\end{equation}
It is clear that for the flat six-dimensional solution with $\sqrt {\det G}=
\sqrt
{\det G}_0$ the massless supersymmetric state of pure $N=4$ supergravity is a
trivial flat space. The non-trivial solutions exist only when there are matter
multiplets. We will find that for our class of solutions not only $N=4$
non-gravitational gauge  multiplets must be present but also  some of them
have to be  non-Abelian to keep supersymmetry preserved with an account taken
of quantum corrections.

The class of massive supersymmetric solutions which we are going to study will
be
closely related to the massless configurations. They will also have the
property that
the mass of the configuration is proportional to the charge of the
four-dimensional
dilaton. Therefore the new mass formulas for the white holes which will be
obtained in this paper will simultaneously  give the  dilaton charge formulas.

\section{Supersymmetric string waves and generalized fundamental strings}

 We have found various examples of  configurations with vanishing
four-dimensional dilaton charge by using some known solutions of the  equations
of motions of effective action of the heterotic string.   Some of them belong
to the
class of exact supersymmetric heterotic string backgrounds and require the
non-Abelian gauge fields to be part of the solution, some other do not seem to
preserve the unbroken supersymmetry when $\alpha'$ corrections are taken into
account. The first class of exact solutions with $SO(8)$ special holonomy is
given  by
supersymmetric string waves (SSW)  \cite{BKO1} and their T-dual
partners, generalized fundamental strings (GFS) \cite{BEK}.    Both the
pp-waves
and fundamental strings  admit a null Killing vector and  belong to the class
of  supersymmetric gravitational waves. The Killing spinor for these solutions
satisfies a null constraint,  and the dimensionally reduced form of these
solutions
always gives electrically charged configurations.

We will consider here the  $u,v$-independent  part of these solutions, which is
described in terms of  {\it nine harmonic functions}, satisfying the flat space
eight-dimensional equation $\sum_{i=1}^8 \partial_i \partial_i \, h (x^i) =0$.
These configurations solve the cohomology constraint ${tr} R^2 - {tr} F^2=0$.

The second class of supersymmetric solutions,  associated with the chiral null
model \cite{HT}, is
described by {\it ten harmonic functions}.  When one of the harmonic functions
is
taken to be a constant, the chiral null model is reduced either to SSW
solutions or
to GFS solutions.  Therefore only for these solutions the embedding of the spin
connection into the gauge group is possible. This leads to the preservation of
unbroken space-time supersymmetry with an account taken of
$\alpha'$-corrections, as
well as to the left-right  world-sheet supersymmetry. However, when all ten
functions are present in the solutions the status of unbroken supersymmetry in
presence of  $\alpha'$-corrections is not clear.  The holonomy of torsionful
spin
connections of this theory, related to the properties of $\alpha'$-corrections,
is a
subgroup  of the non-compact SO(1.9) Lorentz group. This was established in
\cite{KOexact}  for uplifted electrically charged $a=1$ dilaton black holes
(which
form a particular case of the chiral null model) and for the complete chiral
null
model in  \cite{HT}. The spin embedding into the gauge group  is the only known
way to  preserve unbroken supersymmetry. It does not work for generic chiral
null
model since the gauge group of the heterotic string is compact.

All solutions described above admit the null Killing vector and therefore may
be
called gravitational waves. Dimensional reduction of these solutions was
performed
in \cite{KB}.

Here we would like to describe first the solutions which solve the cohomology
constraint $dH=0$ and which   remain supersymmetric even with an  account taken
of
$\alpha'$-corrections.  After this is done we will look for the massive
four-dimensional supersymmetric black holes and study how they approach the
massless states. For this
purpose we start with anomaly free  ten-dimensional solutions of $N=1$
supergravity coupled to supersymmetric Yang-Mills theory. The exact SSW as well
as exact GFS solutions admit  a null Killing vector $l^\mu$ with
$l^2=0$.
This Killing vector generates an isometry in the $v$ direction where
we use light-cone coordinates $x^\mu = (u,v,x^i)$,\  $i=1,\dots ,8$.
The solution consists of the ten-dimensional metric, dilaton, two-form and the
non-Abelian gauge fields.
 The metric and the 2-form field  are both described in terms of the dilaton
$e^{-2\hat \phi (x^i)}$ and one vector function $A_\mu(x^i)$
of the transverse coordinates $x^i$,
\begin{equation}
A_\mu(x^i) = \Bigl\{A_u(x^i) \equiv  - {K(x^i)\over 2},\, A_v=0,\,
A_i(x^i)\Bigr\}     \ .
\end{equation}

For SSW the dilaton has to be taken constant,  $e^{-2\hat \phi(x^i)}  =
e^{-2\hat
\phi_o} $, for GFS the function $K(x^i)$   has  to be a constant,
$K(x^i) = K_0$. Under these conditions both solutions can be described as
follows\footnote{ In the chiral null model \cite{HT}   $e^{2\hat \phi} =F$.}:
\begin{equation}
ds^2 =  2 e^{2\hat \phi} du (dv  + A_{\mu} dx^\mu )- \sum_1^8dx^i dx^i,
\qquad A_v =0 \ ,
\label{wave}\end{equation}
\begin{equation}
B = 2 e^{2\hat \phi} du \wedge (dv + A_{\mu} dx^\mu) \ .
\end{equation}
The non-Abelian gauge field  $V_\mu{}^{IJ}$ is obtained by  embedding of the
torsionful spin connection into the gauge group of the heterotic string.
\begin{equation}
\label{eq:embedding}
V_\mu{}^{IJ}
=  l_\mu V^{IJ}\hskip .5truecm \equiv \hskip .5truecm
\Omega_{\mu-}{}^{ab} =e^{2\hat \phi}  l_\mu {\cal A}^{ab} ,
\ \ \ \ \ a,b,I,J=1,\dots ,8\  .
\end{equation}
The Yang-Mills indices are in the adjoint representation of
$SO(8)$.
The equations that the dilaton $e^{-2\hat \phi (x^i)}$, $K(x^i)$ and $A_i(x^i)$
have to satisfy for the configuration to be supersymmetric and solve equations
of
motion are:
\begin{equation}
\label{eq:Lapl}
\triangle e^{-2\hat \phi} =  \triangle K  = 0\ , \qquad
\triangle\partial^{[i}A^{j]}=0 \ , \qquad \partial^{[i}A^{j]} \equiv   {\cal
A}^{ij} \ ,
\end{equation}
where the Laplacian is taken over the transverse directions only ($\triangle
\equiv
\sum_1^8\partial_i \partial_i$). This solution has   $SO(8)$ symmetry.
Obviously, SSW with constant dilaton and GFS with constant $K$ are special
solutions of this system of equations.\footnote{This solution with constant $K
=
K_0$ is equivalent to the one presented in \cite{BEK} after the shift in the
isometry direction $v' = v -{ K_0 u\over 2} $.} However, for any other
solutions when both
the dilaton and the function $K$ are non-constant, the
proof of unbroken supersymmetry is not available. The holonomy of the
generalized connections in this case includes the non-compact subgroup of the
Lorentz group, since  $\Omega_{\mu-}{}^{0i} $   is not vanishing\footnote{There
is
an apparent discrepancy between the statement about the holonomy of generalized
fundamental string solutions in \cite{BEK} and \cite{HT}.   Using
T-duality rotation from the waves, we have found  that the holonomy  of the SSW
as well as of GFS is
$SO(8)$. Meanwhile in \cite{HT} only the holonomy  of SSW is qualified as that
of
$SO(8)$ and the case of GFS is considered as not special.  The analysis shows
that
there is no discrepancy, however. Spin connections are frame dependent: in the
frame used in \cite{HT} one can still find that the curvatures are such that
the
spin embedding into the $SO(8)$ gauge group solves the problem with
$\alpha'$-corrections for GFS configurations.}.  This spin connection cannot be
embedded into $SO(32)$ or $E_8\times E_8$ gauge group of the heterotic string.
However, without such spin embedding the preservation of supersymmetry at the
quantum level is questionable. Therefore in what follows we will consider
only SSW and GFS with non-Abelian fields included as exact
supersymmetric solutions in our class.

We will limit ourselves to the solutions which in four dimensions correspond
only
to static configurations and not to the stationary ones. For this purpose we
take
$A_1=A_2=A_3 =0 $. Our next specification will be to consider  the solutions
which depend only on $x^1, x^2, x^3$. Thus  the $SO(8)$ symmetry is broken. The
only non-vanishing components of the non-Abelian vector field $V_\mu{}^{IJ}$ in
this
case are     $V_\mu{}^{im} $ with $i=1,2,3$ and $m=4,5,6,7,8$.

\section{ Zero mass configurations with asymptotically flat internal  geometry}

The effective action describing the dynamics of the massless fields of
toroidally compactified heterotic string is described by the bosonic part of
the action of $N=4$
supergravity interacting with 22 abelian $N=4$ vector multiplets.
Ten-dimensional supergravity, dimensionally reduced to four dimensions provides
6 of them. The additional 16  are coming from the ten-dimensional vector
multiplets, or from the gauge sector of the heterotic string. The action in the
form given by
 Maharana-Schwarz \cite{MS}
and  Sen \cite{Sen} is
\begin{eqnarray}\label{action}
S &=& {1\over16\pi } \int d^4 x \sqrt{-\det g} \,  \, \Big[ R -2  g^{\mu\nu}
\partial_\mu \phi \partial _\nu\phi +{1\over 8} g^{\mu\nu} Tr(\partial _\mu
{\cal M} L\partial_\nu  {\cal M} L)
\nonumber \\
&& -{1\over 12} e^{-2 \phi} (H_{\mu\nu\rho})^2
 -  e^{-2 \phi} g^{\mu\mu'} g^{\nu\nu'} F^a_{\mu\nu} \, (L {\cal M} L)_{ab}
\, F^b_{\mu'\nu'} \Big] \, ,
\end{eqnarray}
where $\phi$ is the four-dimensional dilaton, and  ${\cal M}$ is the 28$\times
$28 matrix valued scalar field, describing the moduli. Vector fields include
graviphoton as well as vectors from the gauge multiplets. The  28$\times$28
symmetric matrix $L$   with 22 eigenvalues $-1$ and 6 eigenvalues $+1$ defines
the metric in the O$(6, 22)$ space. We use the units where $G_N=1$,  which
should be taken into account when comparing with $G_N=2$ units often used for
identification of string states.

Dimensional reduction of the chiral null model (without Yang-Mills fields) was
performed in \cite{KB}. The supersymmetric four-dimensional solutions of the
field equations following from the action
(\ref{action}) have
 the following metric (in the canonical frame):
\begin{equation}
ds_{\rm can}^2 = e^{2\phi}dt ^2 - e^{-2\phi} d\vec x^2 \ ,
\label{spher}\end{equation}
and the four -dimensional dilaton is
\begin{equation}
e^{-2\phi} = \Bigl(e^{-2\hat \phi} \, K - \sum_{n= 4}^8 (A_n )^2 \Bigr)^{1\over
2}\ .
\label{4dilaton}\end{equation}
Other fields can be also deduced from the ten-dimensional solution by
dimensional reduction.

The massless solution was found by Behrndt \cite{Klaus2} in the framework of
toroidally compactified heterotic string theory, and it was further generalized
in \cite{BHGAUGE}. It has vanishing dilaton
charge and is obtained when the  functions defining the solutions are taken in
the
form \cite{KB}
\begin{equation}
e^{-2\hat \phi}= 1+  {2\tilde m\over r}\ , \qquad K = 1+
\sum_1^s {2\hat m\over r}\ , \qquad A_n =  {2 q_n\over r}\ ,
\qquad n=4, \dots 8, \qquad r ^2 \equiv \vec x^2.
\label{klaus}\end{equation}
This choice corresponds to the asymptotically flat internal space.
The four-dimensional dilaton is given by
\begin{equation}
e^{-2\phi}= \left( 1+ {2(\tilde m + \hat m ) \over r}    - {4(q^2 - \tilde m
\hat
m)\over r^2}
\right)^{1\over
2},
\end{equation}
where   $q^2 \equiv  \sum_{n = 4}^8 (q_{n})^2$.

In terms of the right- and left-handed charges the dilaton is given by
\begin{equation}
e^{-2\phi}= \left( 1+ {2\sqrt 2 |Q_R| \over r}    - {2 (|Q_L|^2 - |Q_R|^2)\over
r^2}
\right)^{1\over
2}.
\end{equation}
Right-handed charge is the charge   corresponding to  the gravi-photon, and
left-handed charge is the
charge   corresponding to   the vector fields of the matter multiplets. There
is only one
possibility to make these solutions massless: to take
 $\sqrt 2 |Q_R| = \hat m + \tilde m=0 $. The dilaton becomes
\begin{equation}\label{DILATON}
e^{-2\phi}= \left(1-  {4 (\tilde m ^2 +   q^2)  \over
r^2}
\right)^{1\over 2} =   \left(1-  {2|Q_L|^2  \over
r^2}
\right)^{1\over 2} .
\end{equation}
Using the fact that for toroidally compactified string supersymmetric
configurations  $|Q_L|^2 - |Q_R|^2 = - 2 (N_L -1)$, one can see that for
vanishing $Q_R$ the state is characterized by $N_L=0$. Here $N_L$ is a
non-negative integer, describing the total oscillator contribution to the
squared mass
of a state from the left moving oscillators of the string.
Rescaling this solution
for arbitrary value of the dilaton at infinity, $e^{2\phi_0} \equiv g^2$, we
get
finally  the canonical four-dimensional metric in the form \cite{BHGAUGE}
 \begin{equation}\label{DILATON2}
ds_{\rm can}^{ 2} ({\rm el})= \left(1-  {4 g^2   \over
r^2}
\right)^{-{1\over 2}} dt ^2 - \left(1-  {4  g^2   \over
r^2}
\right)^{1\over 2}  d\vec x ^2
 \ .
\end{equation}
The singularity of this configuration at $r=2g$ was found in \cite{BHGAUGE} to
be
a true singularity since the scalar curvature is given by
\begin{equation}
R_{\rm can}^{\rm el} = {4g^2 (2 g^2 + r^2) \over  r (r^2 - 4 g^2)^{5\over 2}} \
{}.
\end{equation}
The origin of this singularity can be traced back to the fact that the volume
of the compactified six-dimensional space shrinks to zero for $r=2g$ since
\begin{equation}
\det G = e^{-4\phi} e^{ 4\hat \phi} ={ r^2 -  4 g^2  \over (r+  2\tilde m)^2} \
{}.
\end{equation}
In the magnetic case   $g\rightarrow {1\over g}$,  and the volume of the
compactified space is
\begin{equation}
\det G = e^{-4\phi} e^{ 4\hat \phi} ={ r^2 g^2-  4   \over g^2(r+  2\tilde
m)^2}
\ .
\end{equation}
It  shrinks to zero at the point $r= {2\over g}$ where the four-dimensional
magnetic solution is singular,
 \begin{equation}\label{DILATON3}
ds^2_{\rm can} ({\rm m}) = \left(1-  {4  \over
g^2 r^2}
\right)^{-{1\over 2}} dt ^2 - \left(1-  {4  \over
 g^2 r^2}
\right)^{1\over 2}    d\vec x^2
 \ .
\end{equation}

All this analysis is valid in the framework of the toroidally compactified
heterotic string with only Abelian vector fields in the solutions,  and   when
one ignores the issue of $\alpha'$-corrections to supersymmetry. From now on we
will
consider only exactly supersymmetric SSW and GFS supplemented by the
non-Abelian fields  and described in Sec. 2.

This means that when both $\tilde m$ and $\hat m$ are non-vanishing,
supersymmetry is anomalous. If only one of these two numbers  vanishes,
which is acceptable from the point of view of the exactness of the solution, we
do
not approach the massless state. Only for $\hat m = \tilde m=0 $ we  can   have
a massless solution  without anomalies. This special
configuration has the dilaton field given by
\begin{equation}\label{DILATON)}
e^{-2\phi}= \left(1-  {4   q^2 \over
r^2}
\right)^{1\over 2}  .
\end{equation}

The presence of the Yang-Mills vector fields in the ten-dimensional solution
will add
some non-Abelian vectors as well as scalars to the four-dimensional solutions.
It
is quite remarkable that for this solution   to be non-trivial the presence of
the
non-Abelian vector field is necessary. In fact, using eq. (\ref{eq:embedding})
we
will find that for the configuration given in (\ref{klaus}) the non-Abelian
part of
the solution in ten  dimensions is given by \cite{BEK}
\begin{equation}
\label{YM}
V_\mu{}^{[in]}
 = l_\mu \, \Bigl(1+  {2\tilde m\over r}\Bigr)^{-1} \;  {x^i q^n \over |x|^3},
\ \ \ \ \ \ i=1,2,3,\ \ \;   n=4,5,6,7,8.
\end{equation}
For the massless configuration presented above with $\hat m = \tilde m=0$ the
Yang-Mills field is
\begin{equation}
\label{YMI}
V_\mu{}^{[in]}
 = l_\mu  \;  {x^i q^n \over |x|^3},
\ \ \ \ \ \ i=1,2,3,\ \ \;   n=4,5,6,7,8.
\end{equation}

The ten-dimensional manifold  in the process of compactification is split into
the four-dimensional manifold $M^4$ with coordinates  $ v=t, x^1,x^2,x^3 $ and
a six-dimensional  manifold $M^6$ with coordinates $ x^4, \dots , x^8$, $u=x^9
$
\cite{KB}.
The moduli space  of our configuration is defined by the six-dimensional metric
$G_{rs}, \; r=4, \dots , 9$,  and by the six-dimensional matrix $B_{rs}$:
\begin{equation}
G_{rs} = \pmatrix{
-  \delta_{mn} &  & e^{2\hat\phi} A_n  \cr
 &   &  \cr
e^{2\hat\phi} A_m & &  -e^{2\hat\phi}  K &  \cr
},   \qquad
B_{rs} = \pmatrix{
0 &  & e^{2\hat\phi} A_n  \cr
 &   &  \cr
- e^{2\hat\phi} A_m  & & 0&  \cr
} .
\label{mod}\end{equation}
In addition, there are non-Abelian vectors and scalars which will come from the
ten-dimensional Yang-Mills field (\ref{YM}). The two matrices $G$ and $B$
together  form the $O(6,6)$ matrix ${\cal M}$ \cite{MS} which appears in the
Bogomolny bound \cite{Sen}.  The ansatz (\ref{klaus}) used in eq. (\ref{mod})
has
the following properties:

\noindent i) The $O (6,6)$ matrix ${\cal M}$ build out of $G$ and $B$ at
infinity
($r\rightarrow
\infty$) is
\begin{equation}
{\cal M}_0 = \pmatrix{
-I & 0 \cr
0 & -I \cr
} .
\label{diag}\end{equation}
 ii) Given this asymptotic value of the matrix ${\cal M}$, there is only one
solution for the massless state which is not changed by quantum corrections:
\begin{equation}
G_{rs} = \pmatrix{
-  \delta_{mn} &  &  {2q_n\over r}  \cr
 &   &  \cr
{2q_m\over r}  & &  -1 &  \cr
},   \qquad
B_{rs} = \pmatrix{
0 &  & {2q_n\over r}   \cr
 &   &  \cr
- {2q_n\over r}   & & 0&  \cr
},  \qquad e^{-2\phi}= \left(1-   {4   q^2 \over
r^2}
\right)^{1\over 2}.
\label{mod1}\end{equation}
 The metric of this configuration   coincides with the metric (\ref{DILATON2}).
Besides, there are Abelian vector fields and non-Abelian vectors and scalars.
The most unusual property of this solution is that the massless state is
described by a static configuration.  We will return to this issue later after
we will find more general exact  massless and massive solutions.

\section{Special  points in the moduli space}

One can find a  solution describing a more general family of BPS-states with a
vanishing ADM mass. For this purpose we may use the fact that in gravitational
wave
solutions in $d=10$ one can use  more general harmonic functions. For the
one-black-hole case one can take
\begin{equation}
e^{-2\hat \phi}=e^{-2\hat \phi_0} +  {2\tilde m \over r}\ ,
\qquad K = K_0 +
 {2\hat m\over r}\ , \qquad A_n = (A_n)_0 + {2 q_n \over r}\ ,
\qquad n=4, \dots 9.
\end{equation}
However, for the solution to be exact we have one constraint
\begin{equation}
\hat m \tilde m =0 \ .
\end{equation}
Indeed, this means that either  $\tilde m  =0$ and the dilaton is constant or $
\hat m =0$ and $K$ is constant, which are the conditions for exactness.
 The four-dimensional dilaton is now given by
\begin{equation}
e^{-2\phi}= \left(\Bigl(e^{-2\hat \phi_0} +  {2\tilde m\over r}\Bigr)\Bigl(K_0
+
 {2\hat m \over r}\Bigr) -\left[(A_n)_0 + {2  q_n \over r}\right ]^2
\right)^{1\over
2} \ .
\end{equation}
This expression can be reorganized as follows:
\begin{equation}
e^{-2\phi}=e^{-2\phi_0}\left(1+   {4M\over r}    - {4  g^2 q^2 \over r^2}
\right)^{1\over
2} ,
\end{equation}
where
\begin{equation}
e^{-2\phi_0}= \left[e^{-2\hat \phi_0}  K_0 - (A_n)_0^2\right]^{1/2}  \equiv
{1\over g^2},
\end{equation}
The mass formula for the  exact SSW case ($\tilde m=0$) is
\begin{equation}
M = {g^2   \over 2}  \left[   e^{-2\hat \phi_0}\, \hat m  - 2 (A_n)_0\, q_n
\right]
\geq 0 \ ,
\label{MSSW}\end{equation}
 whereas  the mass formula for the case of the exact GFS ($\hat m=0$) is
\begin{equation}
M = {g^2 \over 2} \left[{ K_0\, \tilde m   - 2 (A_n)_0\, q_n }\right]  \geq 0 \
{}.
\label{MGFS}\end{equation}

The mass has to be non-negative due to supersymmetric positivity bound, but the
vanishing value of the mass $M$ is not forbidden by supersymmetry.

Thus the metric of the exact non-Abelian electrically charged black hole is
given
by
\begin{equation}
ds_{\rm can}^2 = \left(1+   {4M\over r}    - {4   g^2q^2 \over r^2}
\right)^{-{1\over
2} } dt ^2 - \left(1+   {4M\over r}    - {4  g^2 q^2 \over r^2}
\right)^{1\over
2}   d\vec x ^2 \ .
\label{II}\end{equation}
The moduli space is presented by the matrix ${\cal M}$ which asymptotically
(in the limit $r \to \infty$) is
described in terms of asymptotic values of the matrices $G$ and $B$,
\begin{equation}
(G_{rs})_0 = \pmatrix{
-  \delta_{mn} &  & (A_n)_0\ \cr
 &   &  \cr
 (A_n)_0  & &  -e^{2\hat \phi_0}  K_0 &  \cr
},   \qquad
(B_{rs})_0 = \pmatrix{
0 &  &  (A_n)_0  \cr
 &   &  \cr
-  (A_n)_0   & & 0&  \cr }   .
\label{mod2}\end{equation}
Thus the asymptotic value of the matrix ${\cal M}$ is very different from the
simple diagonal form  in eq. (\ref{diag}). A complete expression for these
matrices
is
\begin{equation}
G_{rs} = \pmatrix{
-  \delta_{mn} &  & (A_n)_0+  {2q_n\over r}  \cr
 &   &  \cr
 (A_n)_0+ {2q_m\over r}  & &  -e^{2\hat \phi}  K &  \cr
},  \qquad
B_{rs} = \pmatrix{
0 &  &  (A_n)_0+{2q_n\over r}   \cr
 &   &  \cr
- ( (A_n)_0+{2q_n\over r})   & & 0&  \cr
},
\label{mod2a}\end{equation}
where for SSW and for GFS we have respectively
\begin{eqnarray}
e^{2\hat \phi}  K&=& e^{2\hat \phi_0} \Bigl(K_0 +
 {2\hat m\over r}\Bigr) \ ,  ~~~~ \tilde m = 0,  \nonumber\\
 \nonumber\\
e^{-2\hat \phi}  K&=&\Bigl( e^{-2\hat \phi_0} + {2\tilde m \over r}\Bigr)^{-1}
K_0
\ ,  ~~~~ \hat m = 0.
\end{eqnarray}
The non-Abelian fields for both configurations in the four-dimensional form can
be deduced from the ten-dimensional form (\ref{YM}).

The moduli space is rather involved and allows to approach the critical points
of
the  massless configuration continuously when the right-hand side in equations
(\ref{MSSW}) and (\ref{MGFS}) tends to zero.

In all cases considered the four-dimensional configuration has a new
singularity
which was not present in the ten-dimensional case. This singularity is present
in the
compactified solution when the volume of the compactified space shrinks to
zero.
For the solutions described above we have
\begin{equation}\label{orbit0}
\det G =e^{-4 \phi}  e^{4 \hat \phi}   = g^4  \left(1+   {4M\over r}   - {4
g^2q^2
\over r^2}
\right) \left(e^{-2\hat \phi_0} +  {2\tilde m \over r}\right)^{-2}  \ .
\end{equation}
At the singularity point
\begin{equation}\label{orbit01}
r_0 =  2(\sqrt{M^2+g^2 q^2}- M) \
\end{equation}
the determinant of the metric of the compactified six-dimensional space
vanishes,
\begin{equation}\label{orbit02}
\det G(r_0) =0 \ .
\end{equation}
In stringy frame the geometry of the electric configuration (\ref{II}) is given
by
\begin{equation}
ds_{\rm str}^2=  \left(1+   {4M\over r}    - {4   g^2q^2 \over r^2}\right)^{-1
} dt ^2 -
  d\vec
x ^2 \ .
\label{stringyelectric}\end{equation}
Note that metric in stringy frame is well defined even in the region $r < r_0$,
where $1+   {4M\over r}    - {4   g^2q^2 \over r^2} < 0$. It may seem
meaningless to continue metric to the region $r< r_0$, since the singularity at
$r= r_0$
is a real curvature singularity. However,
it may be important to have such a continuation in order to investigate the
possibility of
tunneling through the singularity, see Sec. 5. One  may suggest  the following
continuation of the canonical metric (\ref{II}):
\begin{equation}
g_{\mu\nu}^{\rm can} = g_{\mu\nu}^{\rm str}\ \sqrt{\Bigl\vert1+
{4M\over r}    - {4   g^2q^2
\over r^2}\Bigr\vert}
\ ,
\label{stringycanonical}\end{equation}
which gives the following generalization of the canonical metric (\ref{II}):
\begin{equation}
ds_{\rm can}^2 = \left(1+   {4M\over r}    - {4   g^2q^2 \over r^2}
\right)^{-{1} } \left(\Bigl\vert 1+   {4M\over r}    - {4  g^2 q^2 \over
r^2}\Bigr\vert
\right)^{1\over
2}dt ^2 - \left(\Bigl\vert 1+   {4M\over r}    - {4  g^2 q^2 \over
r^2}\Bigr\vert\right)^{1\over 2}   d\vec x ^2
\ .
\label{III}\end{equation}
This continuation preserves an important property of metric
(\ref{stringyelectric}): the determinant of metric changes its sign at $r =
r_0$.

 \section{Exact supersymmetric black holes are white}

It is very tempting to associate singular spherically symmetric
configurations    (\ref{II}), (\ref{III})  with black holes.
However, it would not be   quite correct. Black holes got their name for the
reason
that they strongly attract all particles,  so that even light cannot escape
from a black hole. Gravitational
attraction can be described by the Newtonian potential
$\Phi  = {1\over 2}(g_{00} - 1)$. This yields the usual Newtonian attractive
potential  $\Phi = -{ M\over r}
$ at a large distance from a   massive  Schwarzschild black hole. Meanwhile,
the potential corresponding to the metric $g_{00} =
g^{-1}_{rr} = \Bigl(1 +{4M\over r} - {4q^2\over r^2}\Bigr)^{-1/2}$  at
large $r$ is given by  $\Phi = -{
M\over r} + {q^2\over  r^2}$, and the strength of the gravitational field  is
proportional to
$\Phi' =  { M\over r^2} - {2 q^2\over r^3}$. (For notational simplicity we take
here the coupling constant $g^2 = 1$.) Thus, in the limit $r \to \infty$ we
still have gravitational attraction, but only for the configurations with
positive
ADM mass $M$. However, there is a stable equilibrium for test particles at
$r_c = {2q^2\over M}$, and there is a gravitational {\it repulsion}
(antigravity) for $r < r_c$. (For massless states the gravitational force is
repulsive at all $r$.) This repulsion becomes infinitely strong near
the singularity, which appears at
$r_0 =  2(\sqrt{M^2+q^2}\,- M)$.
Infalling particles cannot touch the singularity at $r= r_0$ and
become totally reflected.

Indeed, one can write an equation of motion for a test particle of a small
mass $m$ in an  external spherically symmetric background  (\ref{II}), see
\cite{Landau}:
\begin{equation}\label{orbit1}
t = E \int dr {g_{rr}\over \sqrt{g_{00}}}\left({g_{rr}}\, E ^2 - g_{00}\,
{L^2\over r^2} - g_{00}
g_{rr}\, m^2 \right)^{-1/2} \ .
\end{equation}
Here $E $ is the test particle energy at $r \to \infty$, and $L$ is its
angular momentum with respect to the center of our configuration. For $g_{00} =
g^{-1}_{rr} = \Bigl(1 +{4M\over r} - {4q^2\over r^2}\Bigr)^{-1/2}$  eq.
(\ref{orbit1}) reads:
\begin{equation}\label{orbit2}
t = E { \int} {dr\, \Bigl(1 +{4M\over r} - {4q^2\over r^2}\Bigr)
\left({E^2\Bigl(1 +{4M\over r} - {4q^2\over r^2}\Bigr)   -
{L^2\over r^2} - m^2\sqrt{1 +{4M\over r} - {4q^2\over r^2}}}\right)^{-1/2} }\ .
\end{equation}

It is clear from this  equation that test particles with any initial energy $E
$
cannot reach the singularity at $r = r_0$. One can easily show that each  test
particle  within a finite time reaches  some minimal radius    $r_{\rm min}  >
r_0$,
and becomes reflected.  For example, in the case $M = 0$, the singularity is at
$r_0 = 2 |q|$, and all massless   test particles with $L \not = 0$ become
reflected at
$r_{\rm min} =   2 |q|\, \sqrt{1+ {L^2\over 4q^2 E ^2}}\, > r_0$. (For
comparison,   all massless particles with $L < 2ME $ are swallowed by the usual
Schwarzschild black hole.) In the case $M = 0$, $L = 0$ massive test particles
are
reflected at  $r_{\rm min} =  {2 |q|\, \Bigl(1 - {m^4\over E ^4}\Bigr)^{-1/2}}>
r_0$.
The only possible exception is the behavior of massless test particles in the
S-state ($m = L =
0$). In this case one should use quantum mechanical treatment similar to the
one
developed in \cite{HW,HM}.  An investigation of this question indicates
that even in this special case particles are totally reflected.  In this sense
our
solutions describe  white holes  rather than  the black ones.

To verify the last statement and to get an additional insight into the nature
of the
repulsive singularity at $r = r_0$ we will study the wave equation for a
massive scalar field, taking the metric of an electric white hole (repulson) in
the form
which allows continuation to $r < r_0$, see eq. (\ref{III}). For the S-wave,
the scalar field equation $\partial_{\mu}(\sqrt{|g|} g^{\mu\nu}
\partial_{\nu}\phi) = - m^2\phi$ in this metric reads:
\begin{equation}\label{orbit11}
\left(1+{4M\over r} -{4q^2\over r^{2}}\right)\ddot\phi - \phi'' -{2\phi'\over
r} =
-m^2\phi\ .
\end{equation}
The solution of this
equation in the WKB approximation for $m \not = 0$ reproduces our previous
results about total reflection, being strongly suppressed at  $r < r_{\rm
min}$.
However,   exact solutions of this equation both for $m \not = 0$ and for $m =
0$ do not vanish and do not exhibit
any kind of singular behavior at the  point $r = r_0$.  To  give a particular
example, one may consider this equation for $M = 0$, $m = 0$.   In this case
equation (\ref{orbit11}) has a stationary solution
$\phi = e^{-iEt}\chi$ in terms of Bessel functions,
\begin{equation}\label{orbit111}
\chi = r^{-1/2}\, J_\nu(Er), \ \ \ \nu_\pm = \pm {1\over 2}\sqrt{1 +4q^2E^2} \
{}.
\end{equation}
These solutions behave in a regular way at $r = r_0$. This suggests that the
singularity at
$r= r_0$ at the quantum level is transparent for massless particles in the
S-wave.
Only one of these two functions, the Bessel function with
$\nu_+ = +
{1\over 2}\sqrt{1 + 4q^2 E^2}$, is normalizable.  It decreases near the
singularity at
$r = 0$ as $r^{\nu_+-1/2}$.  In such a situation the probability flux near the
singularity $r = 0$ vanishes, which shows that even the massless particles in
the
S-wave ($m = 0$, $L = 0$) are totally reflected, though not by the singularity
at $r= r_0$ but by
the singularity at $r = 0$.

Here one should make a cautionary   note. In general,  test particles may
influence
the background. For $M \not = 0$ this does not lead to any problems: one may
consider test particles with energy
$E \ll M$, in which case their influence on the white hole
background can be neglected. Therefore massive states described in our paper do
exhibit the antigravity regime and can be called white holes, or repulsons.

On the other hand, in the limit $M = 0$ our semiclassical
considerations may become somewhat misleading.
Indeed, gravitational repulsion  changes  momentum of a test particle.
This may happen only if the white hole itself  acquires the same momentum with
an opposite sign, which would imply that the massless white hole should start
moving with the speed of light. This may suggest that the  state corresponding
to
a massless white hole  at rest is unstable with respect to infinitesimally
small
external fluctuations, and therefore generically such  states  should be
described as particles (waves) moving with the speed of light
\cite{HullT,Susskind}.

However, it might be impossible to give any boost to a massless white hole
without either forming a bound state with it or making it massive. Indeed,
these
states can be considered massless only at an infinitely large distance from
them, but in this case they do not interact at all. Repulsive force
$-{2q^2\over
r^3}$, which appears at a finite distance from the  center of a massless white
hole, may be interpreted as a gravitational interaction with its massive core.
Thus, gravitational interaction  occurs only with an internal  part of the
massless white hole, which leads to its deformation. Such a deformation may
change energy and the effective mass of the white hole, and then it will be
able to carry
finite momentum without being accelerated to the speed of light.

Another problem appears when one tries to understand the nature  of the
repulsive gravitational field.  The formal  reason of the repulsion is the
existence
of the non-diagonal terms in the metric of six-dimensional compactified space,
see
(\ref{mod1}). However, it would be nice to have a simple intuitive
4-dimensional picture describing the repulsive force from the phenomenological
point of view.  One way of thinking
about it is that the singularity acts on test particles as a body with a
negative
mass. This mass becomes ``screened'' by positive energy density of physical
fields.
Therefore the absolute value of the effective gravitating mass of a sphere of a
radius $r$ decreases at large
$r$ as $-{M^2\over r}$, and finally the total mass vanishes in the limit $r
\to \infty$.

This intuitive picture  is, in fact,  rather counterintuitive. The
states with negative energy do exist in general relativity. For example, the
total
energy of a closed universe is equal to zero as a result of exact cancellation
of positive energy of matter and negative energy of gravitational field. Still
it is hard to imagine how massive or  massless white holes with a repulsive
core could be created
in the process of gravitational collapse of normal matter with positive energy
density.  This could make such solutions very suspicious.  One should note,
however, that the same is
true  for the usual charged stringy black   holes as well: Typically such black
holes cannot be formed in the process of gravitational collapse of   charged
elementary particles. Indeed, in most cases there are no such charged particles
in the
underlying Lagrangian. The description of charged stringy black holes is
somewhat
unconventional as compared with the ordinary Reissner-Nordstrom black holes
containing charged elementary particles. One may consider a sourceless flux of
electric or magnetic field, and then imagine a situation where the
gravitational
force squeezes the flux into a singularity. Then the singularity will look like
an
electrically or magnetically charged particle. To describe such a situation at
a
more formal level, one should  find  the flux of electric, magnetic and
gravitational fields at infinity, and then find an extremum of action with
these
boundary conditions, but without imposing any   boundary conditions
and  solving Lagrange equations   at   the singularity. In particular, there is
no
requirement that the effective charge of the singularity is
carried by an elementary particle, or that the  singularity looks like a normal
particle
with a positive mass. Such requirements appear only if one imposes an
additional
condition that the black hole is formed as a result of gravitational collapse
of
elementary particles. For the reason discussed above, this condition does not
necessarily apply to charged stringy black holes. The best constraint which one
can obtain on
the black hole mass is the supersymmetric Bogomolny positivity bound. This
constraint  applies not to the ``effective mass'' of the singularity, but to
the total
ADM mass, and it is satisfied by the massless and massive white holes
considered
in this paper.

\section{Discussion}

In the previous papers \cite{Klaus2} and \cite{BHGAUGE} the massless $N_L
=0$ states of the toroidally compactified heterotic string have been found.
Those states saturate the supersymmetric positivity bound.
In this paper we have found exact supersymmetric electrically charged
four-dimensional  configurations  whose ADM mass  can  vanish
without the solutions being trivial. One of the features  of these solutions is
the
necessary presence of non-Abelian vectors  and scalars besides the metric,
Abelian vectors and scalars. These   solutions have been obtained as  classical
solutions of the effective ten-dimensional action of the heterotic string
theory.

The
configurations which we have discussed here cannot be associated with the
toroidally compactified string with $O(6,22)$ duality symmetry. The presence of
Yang-Mills fields  required by preservation of supersymmetry at the
quantum level means that  we have only
$O(6,6)$ symmetry with 6 Abelian gravi-photons and 6 Abelian vector multiplets.
But instead of the 16 additional Abelian  vector multiplets, which are
extending
the symmetry of the toroidally compactified string from 6 to 22, we have
non-Abelian vector multiplets.

Therefore
the interpretation of these new configurations  as the states of the properly
quantized string   still has to be investigated.  The quantization conditions
used
for toroidally compactified string should be generalized for the presence of
non-Abelian vector multiplets.

The metric of  the exact supersymmetric configurations which we have studied
has the following general form:
 \begin{equation}
ds_{\rm can}^2 =  \left( 1+ {2\sqrt 2 |Q_R| \over r}    - {2 (|Q_L|^2 -
|Q_R|^2)\over
r^2}
\right)^{-{1\over
2}} dt^2 -  \left( 1+ {2\sqrt 2 |Q_R| \over r}    - {2 (|Q_L|^2 - |Q_R|^2)\over
r^2}
\right)^{1\over
2} dx^2 \ .
\end{equation}
Here $2 (|Q_L|^2 - |Q_R|^2)=q_n^2$,
and the Yang-Mills field is a necessary part of the solutions when $q_n \neq
0$.
The  massive
$a= \sqrt 3$ electrically charged black hole  of Gibbons and Perry \cite{GP} is
included into this class. In fact it is the only case with $ |Q_L|^2 -
|Q_R|^2=2 q_n^2=q_n = 0, \; V^{YM} =0$
for which the configuration does not need the presence of non-Abelian vector
fields   to be exact:  quantum  corrections vanish due to null properties of
the curvature of
the pp-waves with $K= 1+ {M\over r}$ and $e^{-2\hat \phi}=1, A_n =0$
\cite{BKO1}.
This is a
one-parameter extreme black hole solution:
 \begin{equation}
 ds^2 = \left( 1+ {4M \over r}
\right)^{-{1\over
2}} dt^2 -  \left( 1+ {4M \over r}
\right)^{1\over
2} dx^2 \ .
\end{equation}
When the mass of this solution tends to zero, it becomes  trivial, and one half
of unbroken supersymmetry gets restored to the
 completely unbroken supersymmetry of the flat space.
Apart from this massive black hole solution, every other one has
 \begin{equation}
 |Q_L|^2 - |Q_R|^2=2 q_n^2 > 0 \ .
\label{repul} \end{equation}
This means that any solution in this group can become massless ($Q_R =0$)  and
still the metric and the right-handed Abelian vectors (from the gravitational
supermultiplet)
will have some non-vanishing $1/r^2$ terms.  The left-handed Abelian vectors
(from the  non-gravitational vector supermultiplet) as well as  the Yang-Mills
fields  are also present since $ |Q_L|^2 = 2 q_n^2>0$.

Thus  the non-trivial supersymmetric massless configurations described in this
paper    do not exist without the non-Abelian multiplets. Even in the limit of
the vanishing mass one half of the supersymmetry is unbroken  and the
other half is broken and serves to form the supercharge of the ultra-short
multiplet.

A very unusual property of the new set of exact supersymmetric solutions
described in this paper  is the presence of a repulsive singularity when the
condition  $|Q_L|^2 - |Q_R|^2=2 q_n^2 > 0$ is valid. This singularity
appears both for massless ($Q_R = 0$) and for massive ($Q_R  \not = 0$)
solutions. As a result, these solutions can be better classified as  white
holes rather than  the black ones.

Note that the {\it gravitational} repulsion which we are discussing here is
quite different from the repulsive component of interaction between   extreme
black holes, which appears due to the {\it non-gravitational}   interaction  of
their electric, magnetic and dilaton charges \cite{US}. White holes (repulsons)
considered
in this paper repel all particles, either charged or not, with the strength
proportional to their mass. This repulsion, unlike the non-gravitational
repulsion considered in \cite{US}, does not violate the equivalence principle.
If  existence of
repulsons is confirmed, we will have the first realization of  the universal
gravitational  force which repels all  particles  and  therefore can be
associated with antigravity.

One should note, however,
that the interpretation of our solutions as white holes (repulsons) is rather
straightforward for
massive states, but, as we  already mentioned, interaction of particles
with massless states requires a more detailed investigation, and the classical
concept of gravitational repulsion in this case may become inapplicable.
 Formally white holes with vanishing ADM mass are still described by a static
four-dimensional geometry. The limit to the massless state has to be considered
with a special care since normally one would expect that a massless
state has to be described by a wave configuration which admits a null
Killing vector.   However, to have a link to extreme white hole
solutions it is natural to consider those white holes which do not become
trivial
when the mass equals zero. One may try to boost this solution to get the
wave-type
configuration.

Alternatively, after we have found the special points in the moduli space where
the four-dimensional white holes become massless, we may return to the original
form of the ten-dimensional configuration, which from the beginning admitted a
null Killing vector. The simplest one, whose four-dimensional metric is given
in eq. (\ref{DILATON2}), is indeed a supersymmetric pp-wave \cite{BKO1}
described by the
metric, the constant dilaton, the two-form and the Yang-Mills field:
\begin{eqnarray}
ds^2 &=&  2  du  dv  -du^2 + \sum_{n=4} ^8  {2q_{n}\over r} dy^n du -
\sum_{i=1}^3
dx^i dx^i -\sum_{n=4} ^8 dy^ndy^n,
\qquad  e^{2\hat \phi}=1, \nonumber \\
B &=& 2  du \wedge (dv + {2q_{n}\over r}  dx^n),   \qquad
V_u {}^{in}
= {x^i q^n \over  r^3},
\ \ \ \ \ i=1,2,3, \ \ n=4,\dots ,8 .
\end{eqnarray}
The remarkable feature of this solution is the fact that only if the Yang-Mills
field $V$   does not vanish, i.e. $q_{n} \neq 0$, the geometry and the two-form
are
not trivial.  At $q_n =0$ the metric becomes that of the flat space, $ds^2  =
dt^2 -
\sum_{i=1}^9 dx^i dx^i$, and  the three-form $H$  vanishes. The supersymmetry
is not broken at all, it is that of the flat space.
However, as long as $q_{n}\neq 0$, one half of the supersymmetries
is broken, the
condition which the Killing spinor satisfy is  $\gamma^u \epsilon =0$
\cite{BKO1}.

This solution, as well as the more general ones presented in eqs.
(\ref{wave}),
(\ref{eq:embedding}) and described in  Sec. 4,   are exact  solutions with
one-half
of unbroken supersymmetry with an account taken of  perturbative quantum
corrections in
$\alpha'$. The generalization consists in allowing more general asymptotic
values of the ten-dimensional geometry, which is equivalent to allowing the
four-dimensional scalars to have  more general vacuum expectation
values. The new mass formulas for exact supersymmetric non-Abelian white holes
are
presented  in
eqs. (\ref{MSSW}) and (\ref{MGFS}).

Thus we have described the exact supersymmetric non-Abelian configurations
either as
ten-dimensional gravitational waves or as  electrically charged
four-dimensional  white holes, which may be  also called repulsons.
Our exact supersymmetric massless configurations do not exist
without   the  Yang-Mills fields which form a part of the   white hole
configuration.
 One may expect various
 nonperturbative effects including confinement/condensation of
electric/magnetic white holes near the special points of the moduli space where
these solutions become massless. It would be most appropriate to study these
effects, but it is outside of the scope of the present paper. Our main purpose
here was to demonstrate the possibility of the existence of  a new class of
supersymmetric configurations with very unusual properties, and to prepare a
framework
for their subsequent investigation.

\section*{Acknowledgements}

We are grateful  to K. Behrndt, G. Horowitz, A. Sen, L. Susskind, S. Theisen
and A. Tseytlin   for
extremely useful discussions and to the referee for the suggestion to call the
new objects ``repulsons." This
work was  supported
by  NSF grant PHY-8612280.

\end{document}